\documentclass[longbibliography,prl,english,reprint,aps,superscriptaddress]{revtex4-2}

\usepackage{babel,calc,amsmath,amsthm,amssymb,graphicx,subfigure}
\usepackage[T1]{fontenc}
\usepackage{mathdots}
\usepackage[unicode=true]{hyperref}
\usepackage{booktabs}
\usepackage{mathtools}
\usepackage{multirow}
\usepackage{lipsum}
\usepackage[normalem]{ulem}
\usepackage[dvipsnames]{xcolor}
\usepackage{appendix}
\usepackage{tabularx}

\definecolor{linkblue}{rgb}{0.2314, 0.4118, 0.6196}
\hypersetup{
     colorlinks=true,       		% false: boxed links; true: colored links
     linkcolor=linkblue,          	% color of internal links
     citecolor=linkblue,             % color of links to bibliography
     urlcolor=linkblue,          % color of external links
 }

\newcommand{\ket}[1]{\left\vert#1\right\rangle}
\newcommand{\bra}[1]{\left\langle#1\right\vert}

\newcommand{\tr}{\mathrm{Tr}}
\newcommand{\I}{\mathbb{I}}

\newtheorem{result}{Result}

\begin{document}
\renewcommand{\figurename}{Fig.}

%% NAME OVERRIDE-------------------------------------------------------
\newcommand{\extfig}{Extended Data Figure}
\newcommand{\methodsname}{Materials and Methods}
\newcommand{\smname}{Supplementary Material}

\title{%Quantum Contextuality-based high-dimensional quantum key distribution assisted by on-demand single-photon source
%Single-photon quantum key distribution towards semi-device-independent security 
Contextuality-based quantum key distribution with deterministic single-photon sources}
\author{Yu~Meng}
\email{ymeng@dtu.dk}
\affiliation{Center for Hybrid Quantum Networks (Hy-Q), The Niels Bohr Institute, University of Copenhagen, DK-2100 Copenhagen \O, Denmark}
\thanks{Present address: Center for Macroscopic Quantum States (bigQ), Department of Physics, Technical University of Denmark, Fysikvej 307, 2800 Kongens Lyngby, Denmark.}
\author{Debashis Saha}
\affiliation{Department of Physics, School of Basic Sciences, Indian Institute of Technology Bhubaneswar, Odisha 752050, India}
\author{Mikkel Thorbjørn Mikkelsen}
\author{Clara Henke}
\author{Ying Wang}
\affiliation{Center for Hybrid Quantum Networks (Hy-Q), The Niels Bohr Institute, University of Copenhagen, DK-2100 Copenhagen \O, Denmark}
\author{Nikolai Bart}
\author{Arne Ludwig}
\affiliation{Lehrstuhl f{\"u}r Angewandte Festk{\"o}rperphysik,Ruhr-Universit{\"a}t Bochum,Universit{\"a}tsstrasse 150, D-44780 Bochum, Germany }
\author{Peter Lodahl}
\affiliation{Center for Hybrid Quantum Networks (Hy-Q), The Niels Bohr Institute, University of Copenhagen, DK-2100 Copenhagen \O, Denmark}
\author{Adán Cabello}
\affiliation{Departamento de Física Aplicada II, Universidad de Sevilla, E-41012 Sevilla, Spain}
\affiliation{Instituto Carlos I de Física Teórica y Computacional, Universidad de Sevilla, E-41012 Sevilla, Spain}
\author{Leonardo Midolo}
\affiliation{Center for Hybrid Quantum Networks (Hy-Q), The Niels Bohr Institute, University of Copenhagen, DK-2100 Copenhagen \O, Denmark}
%\author{Zheng-Hao~Liu}
%\affiliation{Center for Macroscopic Quantum States (bigQ), Department of Physics, Technical University of Denmark, Fysikvej 307, 2800 Kongens Lyngby, Denmark.}

\date{\today}

%%%%%%%%%%%%%%%%%%%%%%%%%%%%%%%%%%%%%%%%%%%%%%%%%%%%%%%%%%%%%%%%%%%
	
\begin{abstract}
Photons are central to quantum technologies, with photonic qubits offering a promising platform for quantum communication. Semiconductor quantum dots stand out for their ability to generate single photons on demand, a key capability for enabling long-distance quantum networks. In this work, we utilize high-purity single-photon sources based on self-assembled InAs(Ga)As quantum dots as quantum information carriers. %Combined with a witness of quantum contextuality (a more general case of quantum nonlocality), we demonstrated a prepare-and-measure quantum key distribution prototype in free space. Our results do not depend on the assumption that the measurements are ideal or perfectly projective. This work opens a novel perspective for achieving semi-device-independent quantum communication advantages as an intermediate milestone toward full-device independence. 
We demonstrate that such on-demand single photons can generate quantum contextuality. This capability enables a novel protocol for semi-device-independent quantum key distribution over free-space channels. Crucially, our method does not require ideal or perfectly projective measurements, opening a new pathway for robust and practical quantum communication.   
\end{abstract}

%100 character: Photons are central to quantum technologies, with photonic qubits offering strong potential for communication networks. Semiconductor quantum dots can provide on-demand single photons, making them valuable for long-distance quantum communication. Here, we employ high-purity single photons from self-assembled InAs(Ga)As quantum dots as quantum information carriers. We show that these sources can generate quantum contextuality, which we harness to demonstrate a semi-device-independent quantum key distribution protocol in free space. Importantly, our scheme does not rely on ideal projective measurements, pointing toward a robust and practical route to secure quantum communication.
%reviewer: Ana Predojevic ana.predojevic@fysik.su.se
\maketitle

%%%%%%%%%%%%%%%%%%%%%%%%%%%%%%%%%%%%%%%%%%%%%%%%%%%%%%%%%%%%%%%%%%%

%\section{Introduction}

%focus more on the single-photon side, try not mention weak-coherent when do practical key analysis, can add cooperation with BB84 using single photons. 

%%%%%%%%%%%%%%%%%%%%%%%%%%%%%%%%%%%%%%%%%%%%%%%%%%%%%%%%%%%%%%%%%%%

%Among the plethora of practical quantum computing and communication applications, photons stand out as having abundant natural degrees of freedom to encode information that are resilient to environmental noise and are easily manipulated with simple optical operations. 
\textit{Introduction.}---
Photons are well-suited for quantum communication and computation due to their multiple degrees of freedom for encoding information, noise resistance, and ease of manipulation. 
Besides, photons travel at the speed of light, especially lossless at telecom wavelengths in optical fibers, thus making them ideal candidates for transmitting information within distributed quantum networks. One significant application is quantum key distribution (QKD) \cite{gisin2002quantum}, where photonic qubits enable secure communication between two nodes, warranting security against eavesdroppers with assumptions limited by the laws of quantum theory. The most well-known protocol is BB84, which was proposed by Bennett and Brassard \cite{bennett1984quantum},
in which information is encoded into single-photon states and measured by the receiver. While such prepare-and-measure (PAM) schemes are conceptually simple and experimentally accessible, they still face practical vulnerabilities originating from imperfections in channels and devices.

%There, secret keys are generated in a one-way architecture: Alice encodes information on a string of single photon states and sends them to Bob, who then measures them. Such a brief prepare-and-measure (PAM) protocol is easy to implement in practical applications, but it leaves many loopholes in channels and devices that need to be addressed.

%On the source side, most implemented quantum key distribution experiments utilize commercially available weak-coherent lasers as photon sources. However, due to the inherent Poisson distribution of such sources \cite{vajner2022quantum}, multiphoton components cannot be solved theoretically, leaving the risk of photon-number-splitting attacks \cite{lutkenhaus2002quantum}. Alternatively, employing the decoy-state method \cite{wang2005beating} can enhance resistance against such attacks, albeit at the cost of increased experimental complexity. Henceforth, the high-efficiency generation and manipulation of single photons is decisive irrespective of which kind of communication protocols.

Solid-state single-photon emitters \cite{aharonovich2016solid} have proven to be powerful and versatile sources for photonic quantum systems. The discrete energy level due to the strong confinement enables the deterministic generation of single-photon states or even multi-photon entanglement \cite{uppu2020scalable,meng2023deterministic}. Photonic crystal waveguides have been utilized to tailor light–matter interaction in a solid-state environment \cite{lodahl2015interfacing}, achieving near-unity emitter-photon cooperativity, enabling scalable and efficient hardware for quantum computing and networking \cite{uppu2021quantum,chan2025practical}.
%(*** Cite Uppu, Carolan, Nature nano, and the recent paper from Ming Lai and Stefano ***). 
In earlier work, we employed self-assembled InGaAs quantum dots (QDs) in photonic crystal waveguides to generate single-photon states with high source brightness and long-term stability. We demonstrated a BB84 quantum key distribution field trial \cite{zahidy2024quantum}. More recently, Pan and colleagues implemented the protocol with quantum dots embedded in open cavities, achieving performance beyond the fundamental rate limit of weak coherent states\,\cite {zhang2025experimental, ding2025high}. These results collectively highlight the deterministic single-photon source as a promising platform for secure and efficient quantum communication. 

While quantum dot single-photon sources (SPSs) provide remarkable performance, employing them alone does not necessarily guarantee information-theoretic security. The ultimate goal for quantum communications is the device-independent quantum key distribution (DI-QKD), whose security is inherently ensured by quantum nonlocality \cite{acin2007device,kolodynski2020device}. 
%non-locality of quantum mechanics 
%As long as the correlation between two communication parties (Alice and Bob) violates Bell's inequality, the key pairs can be securely established between them without characterizing the measurement apparatus or any other physical devices.
The implementation of DI-QKD protocol requires loophole-free Bell test %{\color{blue} This is not true: the locality loophole can be left open and, instead,  invoke the no-leakage assumption which is standard in cryptography. I would replace ``all loopholes closed'' by ``most loopholes closed''} 
\cite{hensen2015loophole,shalm2015strong,giustina2015significant}, which calls for high demands in practice, i.e., high-quality entanglement preparation among spacelike intervals and near-perfect quantum measurements. 

In this work, we introduce a \textit{semi-device-independent} quantum key distribution protocol that resorts to quantum contextuality \cite{Budroni:2022RMP}, a generalized form of quantum nonlocality, as a security check. Similar to DI-QKD, its security is testified when contextuality inequalities are violated, but with a restriction on the system's dimension, which makes it not a full-DI system. Moreover, such contextuality-based QKD protocols are compatible with the PAM architecture, thus without the need to distribute entanglement states among Alice and Bob. We emphasize that only the single photons allow contextuality-based QKD protocol to fully exploit its semi-device-independent security advantage, and achieve a higher level of secure key rate compared to the weak coherent states. We demonstrate this advantage by using a semiconductor quantum dot as the single-photon emitter and realize a proof-of-principle QKD prototype experiment in free space. 
This work demonstrates the practical benefits of quantum dot single-photon sources in quantum communication and bridging the gap between the conventional single-photon systems and more advanced semi-device-independent QKD protocols, thus a significant step beyond conventional implementations relying on attenuated laser pulses.

%In this work, we report a free-space experimental realization of a prepare-and-measure QKD protocol that integrates quantum contextuality as a built-in security witness, leveraging the deterministic nature of a quantum dot single-photon source. This demonstrates the practical benefits of quantum dot single-photon sources in quantum communication and bridging the gap between conventional SPS-QKD systems and more advanced semi-device-independent QKD protocols.

%Here, we pose the question of whether it is possible to establish a quantumness witness in one-way QKD protocols as is similar to the Bell inequality violation in DI-QKD, but without relying on entanglement, thereby achieving higher security than the BB84 protocol. Our answer is affirmative. We will show in the later section that, by using self-assembled InAs quantum dots in photonic crystal waveguides as single-photon sources, we implemented a quantum contextuality-based QKD prototype in free space and demonstrated its significant enablers in single-photon one-way QKD. 

%%%%%%%%%%%%%%%%%%%%%%%%%%%%%%%%%%%%%%%%%%%%%%%%%%%%%%%%%%%%%%%%%%%

%\section{Single photon source for semi-device-independent QKD }
%\subsection{Quanutm contextuality}

%%%%%%%%%%%%%%%%%%%%%%%%%%%%%%%%%%%%%%%%%%%%%%%%%%%%%%%%%%%%%%%%%%%
\textit{Single photon source for semi-device-independent QKD.}---%We will first explain what \textit{quantum contextuality} is and discuss how it can be implemented to realize a QKD protocol with semi-device-independent security, and finally, emphasize why a single photon source is the best candidate for such protocols. 
Quantum contextuality refers to the fundamental property that the outcome of a measurement on a quantum system cannot be thought of as revealing a pre-existing value independent of which other compatible measurements are jointly measured. 
Similar to the Bell inequality violation, which denies the local hidden variable model, quantum contextuality can be certified via the \textit{noncontextuality (NC) inequalities}, which any noncontextual hidden-variable model must satisfy, but quantum systems with dimension $d \geq 3$ can violate. Such violations reveal the intrinsic unpredictability of quantum systems, thus serving as the fundamental origins of randomness in quantum communication tasks. 
%It can be certified via the \textit{noncontextuality (NC) inequalities}, which any noncontextual hidden-variable model must satisfy, but quantum systems with dimension $d \geq 3$ can violate. 
%Theoretically, a noncontextual hidden variable (NCHV) model cannot describe contextuality correlations. 
It is well-known that the loophole-free Bell tests demonstrate fully device-independent (DI) advantages; their implementations, however, are based on the distribution of quantum entanglement among distant parties. 
In contrast, the experimental verification of the NC inequality can be implemented within a prepare-and-measure (PAM) scenario, but with an additional assumption on the system's dimension. Such a restriction renders them semi-device-independent (SDI) features, as the internal workings of the devices remain uncharacterized — treated as black boxes. 

A recent work by our co-authors demonstrates that any quantum contextual correlation generated by sufficiently small-dimensional quantum systems can exhibit a quantum communication advantage \cite{gupta2023quantum}, when properly designing the PAM communication settings \cite{cabello2016simple, klyachko2008simple,cabello2013simple,cabello2008experimentally}. Its communication security is ensured by the \textit{monogamy of contextuality}\,\cite{ramanathan2012generalized,kurzynski2014fundamental,saha2017activation}. %as long as the contextuality correlation between the two communication parties violates the NC inequality, no third party (e.g., an eavesdropper Eve) can simultaneously share the same amount of correlation with either of them. 
This work presented an application of how to interpret such communication advantage into SDI-QKD tasks. Such QKD protocol frameworks provide a higher level of security than traditional protocols like BB84, which rely primarily on the observed quantum bit error rate without certifying the fundamental source of quantum randomness.

In most practical QKD implementations, weak coherent states from attenuated lasers are the typical photon sources. However, their intrinsic Poissonian distribution inevitably leads to a non-negligible fraction of multi-photon emissions, leaving the system vulnerable to photon-number-splitting attacks and compromising the information-theoretic security. Although employing the decoy-state method\,\cite{wang2005beating} can enhance resistance against such attacks, albeit at the cost of increased experimental complexity.
In addition to these well-known implementation-level issues, the weak coherent laser also poses a fundamental negative effect from the protocol perspective, i.e, its multi-photon components decline the violation of the noncontextuality inequalities \cite{zhang2019experimental}, thereby weakening or even destroying the very security witness that underpins contextuality-based QKD protocols.
%Therefore, to faithfully implement contextuality-based SDI-QKD protocols, the use of deterministic single-photon sources is essential.
Solid-state emitters, such as InAs quantum dots embedded in GaAs, provide high-purity, on-demand single photons with sub-Poissonian statistics that effectively suppress multi-photon components at the source level, and also offer a high violation of the contextuality inequality. This makes deterministic single-photon sources uniquely compatible with the requirements of contextuality-based SDI-QKD and fundamentally superior to weak coherent sources in this context.

%quantum key distribution (QKD)QKD. 
In the next section, we first detail how to implement a contextuality-based QKD protocol in our free-space experiment using quantum dots as single-photon emitters, then compare the violation of the 
Klyachko-Can-Binicio{\u{g}}lu-Shumovsky
(KCBS) inequality \cite{klyachko2008simple} using both weak coherent states and true single-photon sources, to demonstrate the superiority of the latter. Finally, we do the security analysis based on different quantum contextuality correlations to highlight that adopting a single-photon source in quantum communication is not only advantageous but essential for realizing both the security and the quantum advantage offered by contextuality.

%%%%%%%%%%%%%%%%%%%%%%%%%%%%%%%%%%%%%%%%%%%%%%%%%%%%%%%%%%%%%%%%%%%

\begin{figure*}[t]
\centerline{\includegraphics[width=17.2cm]{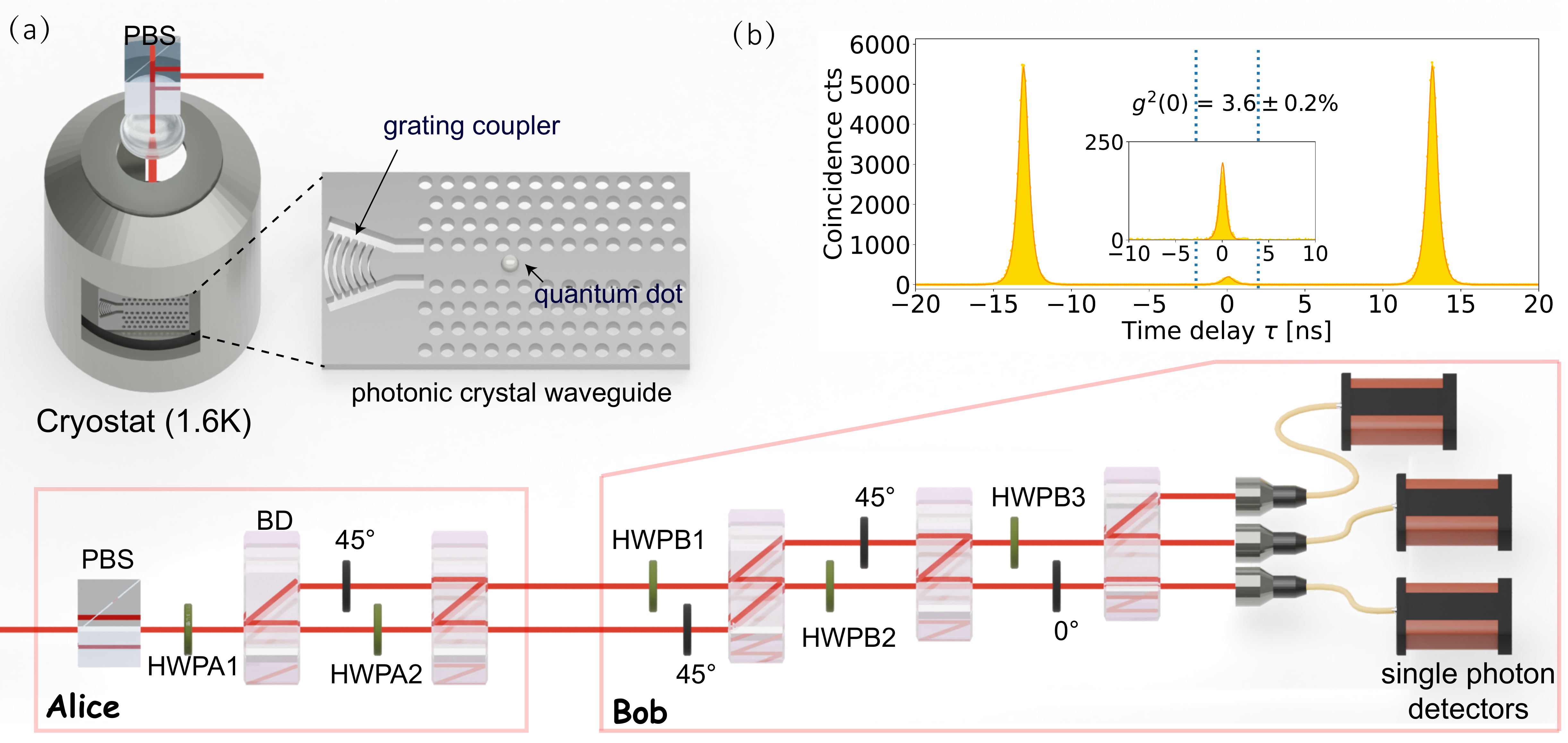}}
\caption{(a) The architecture of contextuality-based QKD protocol prototype in free space using the hybrid of path and polarization encoding. A photonic crystal waveguide collects the emitted photons from the quantum dot and guides the photons to the shallow etched grating out-couplers in the cold chamber (AutoDry 1000).  (b) Experimental result of the HBT experiment: $g^2(0) = 3.6\% \pm 0.2\%. $ It is done by the Hanbury Brown and Twiss (HBT) experiment. }
\label{fig1}
\end{figure*}

%%%%%%%%%%%%%%%%%%%%%%%%%%%%%%%%%%%%%%%%%%%%%%%%%%%%%%%%%%%%%%%%%%%

%%%%%%%%%%%%%%%%%%%%%%%%%%%%%%%%%%%%%%%%%%%%%%%%%%%%%%%%%%%%%%%%%%%

%\section{Quantum contextuality QKD with single-photon source}

%%%%%%%%%%%%%%%%%%%%%%%%%%%%%%%%%%%%%%%%%%%%%%%%%%%%%%%%%%%%%%%%%%%
\textit{Experimental setup.}---
Following the theoretical structure \cite{gupta2023quantum}, the contextuality-based QKD strategy is briefly reviewed as follows: Alice’s preparation device and Bob’s measurement device are treated as black boxes, but with a known dimensional constraint: both are limited to a three-dimensional Hilbert space, effectively realizing qutrit systems. For each round, Alice prepares a quantum state according to randomly chosen inputs $x$, and Bob performs his measurement determined by his inputs $y$, yielding an outcome $b$. After a large number of rounds, Alice publicly announces some random rounds of her choice $x$. Then, combined with his choice of $y$, Bob computes a value of a figure of merit $S$ using the input–output correlations $p(b|x,y)$. As long as $S$ exceeds a classical bound $S^c$ (the maximum attainable under the corresponding noncontextual hidden-variable models), the security is verified. Finally, the secure keys can be sifted from the remaining rounds. We refer the reader to that work\cite{gupta2023quantum} for a full theoretical treatment of the details. %, including the exclusivity graph, preparation and measurement settings, and the figure of merit. 

To meet the requirement that the underlying physical system possesses a Hilbert space of dimension three, we encode the qutrit states using a hybrid of polarization and path degree of freedom of a single photon. 
We use several calcite beam displacers to build two passively phase-stable optical blocks corresponding to Alice’s state preparation and Bob’s measurement. The beam displacers (BDs) transmit vertically (V) polarized photons while displacing horizontally (H) polarized photons, effectively generating distinct spatial modes. Another mode is encoded using the polarization degree of freedom; we use half-wave plates (HWPs) to manipulate the two polarization modes (H/V) on one of the paths while keeping the polarization mode of the other path always H- or V-polarized. As shown in Fig.\,\ref{fig1}(a), neither Alice's preparation device nor Bob's measurement device always consists of one photon with two paths. %Using only a few HWPs and BDs, we can implement arbitrary state preparation and measurement. 
In our experiment, all possible combinations of preparation and measurement settings were manually realized in free space by using HWPs and BDs, which effectively maps out the full specification of the contextuality-based QKD protocol. Unlike a fully operational QKD system that relies on randomized inputs in each round, our setup performs passive preparation and measurement to validate the protocol in a proof-of-principle manner. Details of the implementation are provided in the Supplemental Material \cite{sm}. %As for single-photon source, we utilize the self-assembled InGaAs quantum dot embedded in a photonic crystal waveguide \cite{uppu2020onchip,uppu2020scalable}. Photons are collected through a high-numerical-aperture microscope objective via a grating coupler fabricated at the waveguide output \cite{zhou2018highefficiency}.

To investigate the impact of photon source statistics on contextuality, we measure the violation of the KCBS inequality using both a quantum dot (QD) single-photon source and a weak coherent laser. The InGaAs QD sample is placed in a closed-cycle cryostat (attoDry 1000) at $4\, \rm {K}$. Resonant excitation of a single InGaAs transition at 938 nm is performed using a pulsed mode-locked laser (Picus Q, 80 MHz repetition rate), yielding a detected count rate of approximately 3 MHz. The single-photon source purity is described by the second-order correlation function $g^{(2)}(0)$ using the Hanbury Brown and Twiss (HBT) intensity interferometry. We get $g^{(2)}(0)=3.6\pm 0.2\% $ for our quantum dot candidate in the fiber on avalanche photodiodes (Excelitas SPCM-CD3371H). 

As a benchmark, we repeat the same experimental procedure for getting contextuality correlation using attenuated coherent light generated from a Ti: Sapphire laser (MIRA 900 with Verdi-V8) under various mean photon numbers $\mu$. The weak coherent states follow a Poissonian distribution, and the multi-photon components are tunable via optical attenuation. Here, only registered photon detections are considered; all non-click events are discarded, as they will be treated as channel loss in the real QKD procedure. Additionally, this is based on the assumption of fair sampling, which treats all lost photons as having the same behavior as the registered photons.

%%%%%%%%%%%%%%%%%%%%%%%%%%%%%%%%%%%%%%%%%%%%%%%%%%%%%%%%%%%%%%%%%%%

%\section{Results}
%\subsection{KCBS inequality violation for two sources}

%%%%%%%%%%%%%%%%%%%%%%%%%%%%%%%%%%%%%%%%%%%%%%%%%%%%%%%%%%%%%%%%%%%

\begin{figure}[t]
\includegraphics[width=\columnwidth]{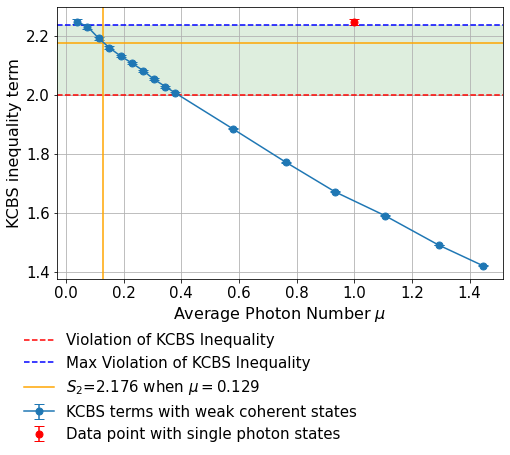}
\caption{Experimental violation of the KCBS inequality using quantum dot single-photon states (red) and weak coherent states (blue) at various average photon numbers $\mu$. For fair comparison, the quantum dot source is assumed to emit exactly one photon per pulse, corresponding to $\mu = 1$. Error bars are estimated using a Monte Carlo method, but are smaller than the marker size and not visible at this scale.}
\label{fig2}
\end{figure}

%%%%%%%%%%%%%%%%%%%%%%%%%%%%%%%%%%%%%%%%%%%%%%%%%%%%%%%%%%%%%%%%%%%
\textit{Results.}---
We here consider a dimension witness for three-dimensional quantum systems as a criterion, as detailedly demonstrated in Sec. I of the Supplemental Material \cite{sm}. %, we consider a dimension witness for three-dimensional quantum systems. 
The witness $S = S_1 + S_2$ consists of two parts: $S_1$, which quantifies the orthogonality between the prepared and measured bases, and $S_2$, which corresponds to the KCBS noncontextuality inequality. They are all obtained within the experimental setup in Fig.\,\ref{fig1}(a) by choosing different combinations between Alice's preparation and Bob's measurement basis, i.e., different combinations of $x$ and $y$. 
The experimental results of $S_2$ between different photon sources are shown in Fig.~\ref{fig2}, and the specific experimental data are listed in the Supplemental Material\,\cite{sm}. For a direct comparison, we assume the ideal quantum dot (QD) emits exactly one and only one photon per pulse, i.e., the average photon number $\mu=1$. With the QD source (red dot), we observe a significant violation of this witness with a measured value of $S_2=2.2463\pm0.0091$, exceeding the classical bound of $2$. For the weak coherent laser (blue dots), we get different average photon numbers $\mu$ by applying different attenuations to the same source. 
While for the orthogonality part, we obtain $S_1=29.8238\pm 0.0129$, this gives us $S=S_1+S_2=32.0701\pm 0.0221$, which exceeds the classical bound of $S^c =32$. It should be noted that the classical bound does not rely on the assumption that measurements are ideal or projective \cite{gupta2023quantum}. 

%With the QD source (red dot), we observe a significant violation of the KCBS inequality with a measured value of $2.2463$, exceeding the classical bound of 2.  In contrast, for the weak coherent source (blue dots), the KCBS inequality decreases as the average photon number increases, illustrating the detrimental effect of multi-photon events in contextuality tests.  Unlike weak coherent sources, the QD single-photon source essentially eliminates multi-photon components. It supports robust key generation with high KCBS inequality violation, thus establishing its strength in semi-device-independent protocols, as we explore next.

%\subsection{SDI-QKD security for the communication tasks}

%The different behavior of the two sources in the KCBS inequality violation raises the question of how they would affect the performance in quantum communication tasks. To address this, we now explore the quantitative link between the contextuality violation and secure key rate in our contextuality-based semi-device-independent QKD protocol.  As demonstrated in Section III. The total contextuality witness $S = S_1 + S_2$ is composed of two parts: $S_1$, which captures the orthogonality between the prepared and measured bases, and $S_2$, which quantifies the violation of the noncontextuality inequality.

%Using our experiment setup in free space, we can obtain all the correlation combinations that contribute to either of them. For the QD source, we examine the experimentally observed values that are $S_1 = 29.8238$ and $S_2 = 2.2463$, such that $S_B = S_1 + S_2 = 32.0701$. 

The semi-device-independent QKD protocol relies on the violation of this witness. Here, the eavesdropper has full control over the preparation and measurement devices, optimizing them to maximize the ability to guess the key, under the constraint that the dimension of the quantum systems is three. Nevertheless, the unknown states and measurements must reproduce the experimentally observed violation, which inherently restricts the extent of possible eavesdropping. An imperfect violation allows limited eavesdropping via cloning attacks, in which the eavesdropper uses an ancilla initial state and performs a controlled unitary gate that clones the communicated state. However, to maintain a sufficiently high violation from the practical experimental result, the eavesdropper’s interference must be minimal, ensuring that the key distribution between Alice and Bob remains effectively secure. We use the semi-definite programming (SDP) relaxation technique based on the Navascués--Pironio--Acín (NPA) hierarchy \cite{npa} and the Navascués--Vértesi hierarchy \cite{NVHierarchy}, which provides an upper bound on Eve’s guessing probability. 
%Under these experimentally realistic parameters,
Under the generic cloning attack, we obtain a positive key rate of 0.1004. For an idealized case where $S_1$ reaches its theoretical maximum, i.e., the orthogonality relations between the prepared and measured basis are perfect, the best eavesdropping strategy for Eve is always to guess the key with the best possible strategy; in this case, the key rate can be as high as 0.174. Details of the key analysis are provided in Sec. II of the Supplemental Material \cite{sm}.

We next apply the same SDP-based analysis to the weak coherent source. As shown in Fig.\ref{fig2}, our experiment only provides $S_2$ for different mean photon numbers $\mu$. We assume $S_1$ to be the same as in the quantum dot case. This assumption is reasonable, as $S_1$ reflects the orthogonality of the preparation and measurement bases, which are implemented identically for both sources in our optical setup, and not the intrinsic properties of the photon source. Moreover, as this approach is semi-device-independent, it relies solely on observed contextuality violations for security guarantees, without assuming detailed knowledge of the source.

\begin{figure}[t]
\includegraphics[width=\columnwidth]{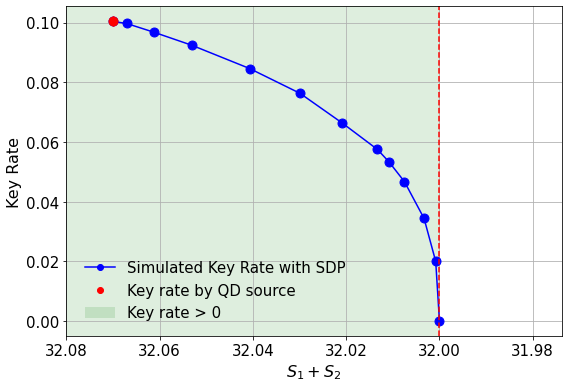}
\caption{Secure key rate analysis for weak-coherent lasers and quantum dot single photon sources with different $S_2$ values, when assuming $S_1$ to be the same as in the single photon source case.}
\label{fig3}
\end{figure}

As shown in Fig.~\ref{fig3}, the optimized secure key rate for the weak coherent states (blue curve) decreases as the KCBS inequality term $S_2$ drops when keeping $S_1$ as the experimentally obtained value. As soon as  $S_2 < 2.1762$, the sum of $S_1+S_2$ becomes too weak to yield a positive key rate. This is only possible when the average photon number $\mu <0.129$, to mostly inhibits the detrimental effect of multi-photon contribution. However, operating at such a low $\mu$ drastically reduces the overall valid detection events. Such an intrinsic trade-off between source brightness and secure key rate fundamentally limits the practicality of weak coherent pulses in the contextuality-based QKD protocols.
In contrast, the deterministic single-photon emission of the quantum dot source supports both strong contextuality violations and robust key generation, even under experimental imperfections.

Beyond its application to QKD, this experiment demonstrates an advancement toward quantum random number generation, offering a novel and impactful approach to the literature on contextuality-inspired quantum randomness \cite{Abbott-pra,Carceller-prl,liu2025randomness,singh2024}. By employing SDP techniques \cite{NVHierarchy}, we estimate a lower bound on the amount of certified randomness within the semi-device-independent framework. In the ideal case, assuming perfect orthogonality and taking the maximum value of $S_1$, we obtain a certified randomness of 0.86 bits from the measured value $S_2$. The details are provided in Supplemental Material \cite{sm}.

%%%%%%%%%%%%%%%%%%%%%%%%%%%%%%%%%%%%%%%%%%%%%%%%%%%%%%%%%%%%%%%%%%%

%%%%%%%%%%%%%%%%%%%%%%%%%%%%%%%%%%%%%%%%%%%%%%%%%%%%%%%%%%%%%%%%%%%

%----------DISCUSSION----------

%\section{conclusion}

%%%%%%%%%%%%%%%%%%%%%%%%%%%%%%%%%%%%%%%%%%%%%%%%%%%%%%%%%%%%%%%%%%%
\textit{Conclusion.}---
In this work, we experimentally realize a prototype of semi-device-independent quantum key distribution %(SDI-QKD) 
based on quantum contextuality using the on-demand, high-purity nature of the InAs/GaAs quantum dots single-photon source.

Our experiment begins with a direct demonstration of the violation of a noncontextuality inequality using single photons, confirming the quantum origin of the generated key. 
We then benchmark the performance of our quantum dot single-photon source against conventional weak coherent states from attenuated lasers. This comparison reveals a fundamental trade-off between brightness and security: while weak coherent states are easier to generate, their inherent multi-photon emissions reduce the contextuality violation and expose the protocol to photon-number-splitting attacks. In contrast, our deterministic quantum dot source combines a near-unity single-photon probability ($\mu \approx 1$) with sub-Poissonian statistics ($g^2(0) \approx 0$), inherently suppressing multi-photon contributions and enhancing contextuality-based security.

These results underscore the crucial role of high-quality single-photon sources in contextuality-based semi-device-independent quantum key distribution protocols. By simultaneously enabling stronger violations of noncontextuality inequalities and delivering higher secure key rates—even under realistic imperfections—our quantum dot source demonstrates a clear advantage over weak coherent states, reinforcing the view that deterministic single-photon emitters are essential for advancing both the foundations and the practical implementation of secure quantum communication.

Since contextuality tests are naturally compatible with the prepare-and-measure scenario, this approach requires no entanglement distribution or loophole-free Bell tests—security is guaranteed solely through the violation of a noncontextuality inequality. This relaxes the resource demands of fully device-independent quantum key distribution %{\color{blue} Some journals, including PRL, prefer avoiding acronyms in the conclusions} 
while still enabling strong security guarantees and higher key rates. 
Overall, our results demonstrate that solid-state quantum dot emitters are not only compatible with but ideally suited for contextuality-based semi-device-independent quantum key distribution. 
This work not only extends the standard prepare-and-measure quantum key distribution framework into high-dimensional quantum communication but also marks a promising step toward practical, device-independent security. Beyond its practical implications, the experiment also offers new insight into the foundational role of contextuality in quantum information science.

%%%%%%%%%%%%%%%%%%%%%%%%%%%%%%%%%%%%%%%%%%%%%%%%%%%%%%%%%%%%%%%%%%%

\noindent\textsf{\textbf{ACKNOWLEDGMENTS}}\\ 
We thank Klaus Mølmer, Armin Tavakoli, and Davide Rusca for heuristic discussions. 
We gratefully acknowledge financial support from the Danish National Research Foundation (Center of Excellence Hy-Q DNRF139), 
Innovationsfonden (No.~9090-00031B, FIRE-Q), 
European Research Council (ERC) under the European Union’s Horizon 2020 research and innovation programme (Grant agreement No.~949043, NANOMEQ),
Styrelsen for Forskning og Innovation (FI) (5072- 00016B QUANTECH),
BMBF QR. N project 16KIS2200, QUANTERA BMBF
EQSOTIC project 16KIS2061, and the DFG excellence cluster ML4Q project EXC 2004/1. DS acknowledges financial support from STARS (STARS/STARS-2/2023-0809), 
Govt. of India. AC was supported by AEI/MICINN (Project No.~PID2020-113738GB-I00),
the Canada-EU project ``Foundations of Quantum Computational Advantage'' (FoQaCiA) (doi: 10.3030/101070558).

%%%%%%%%%%%%%%%%%%%%%%%%%%%%%%%%%%%%%%%%%%%%%%%%%%%%%%%%%%%%%%%%%%%

%--------------------REFS--------------------

\let\oldaddcontentsline\addcontentsline% Store \addcontentsline
\renewcommand{\addcontentsline}[3]{}% Make \addcontentsline a no-op

\bibliographystyle{apsrev4-2}
\bibliography{ref}

\let\addcontentsline\oldaddcontentsline

\onecolumngrid
\vfill

\clearpage
\clearpage
\newpage
\appendix
\onecolumngrid

\setcounter{equation}{0}
\setcounter{figure}{0}
\renewcommand{\theequation}{S\arabic{equation}}
\renewcommand{\thefigure}{S\arabic{figure}}
%\thispagestyle{empty}

% \fi

\begin{center}
\textbf{\large Supplementary Material for ``Contextuality-based quantum key distribution with deterministic single-photon sources''}
\end{center}

\tableofcontents

\section{Experimental implementation}
    
\subsection{Optical Setup Summary}

\begin{figure}[h!]
    \centering
\includegraphics[width=0.77\linewidth]{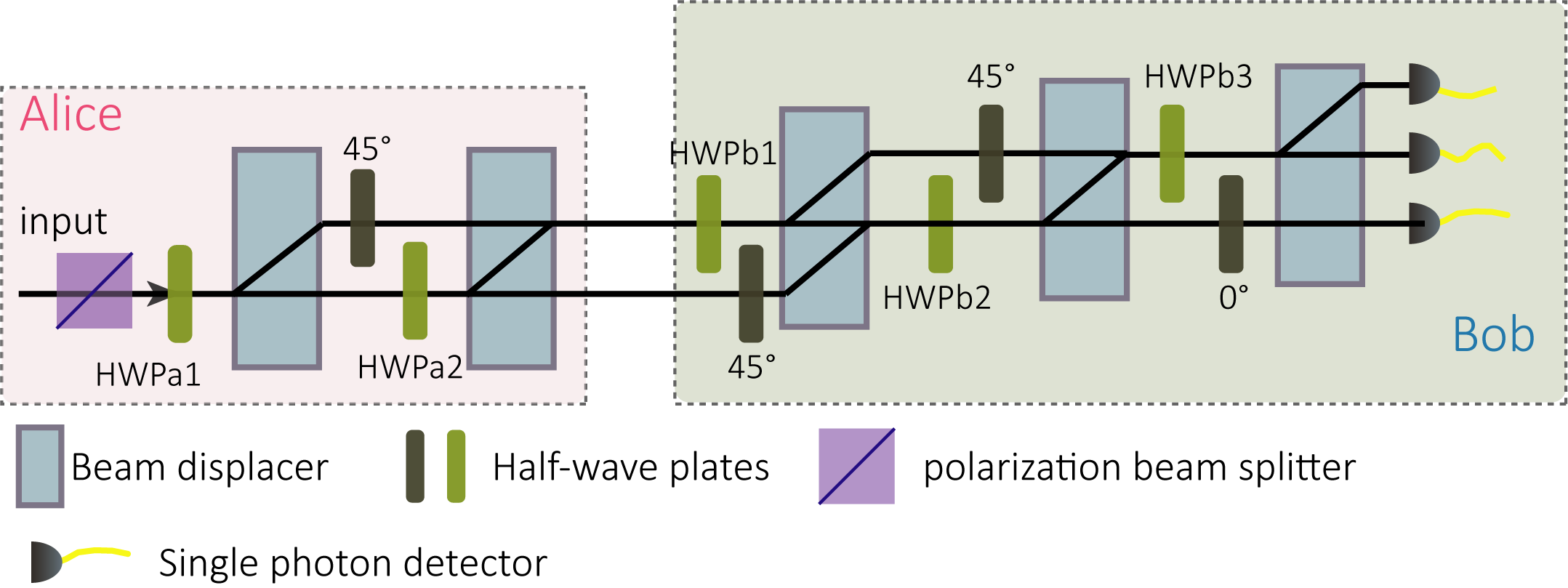}
    \caption{The 2d experimental setup}
    \label{figs1}
\end{figure}
    
In the experiment, we will use the hybrid of polarization and path degrees of freedom of a single photon to encode and prepare arbitrary qutrit states. For manipulating the path degree of freedom, we use beam displacers (BDs), which are birefringence crystals ($\rm YVO_4$) that split an unpolarized light beam into two parallel, orthogonally polarized beams as $|U\rangle$ (up) and $|D\rangle$ (down). BD transmits vertically polarized photons while displacing horizontally polarized photons. Then, the half-wave plates (HWPs) are used to manipulate the ratios on different paths. So that the three modes constituting a qutrit are associated with the horizontal polarization in the upper mode, the vertical polarization in the upper mode, and the horizontal polarization in the lower mode, i.e., $\{|0\rangle=|U H\rangle,|1\rangle=|U V\rangle,|2\rangle=|LH\rangle\}$, where $U (L)$ denotes the upper (lower) path of single photons in the beam displacers, and $|H(V)\rangle$ denotes their horizontal (vertical) polarizations. In the lower path, the photon is always polarized horizontally in some places and vertically in others. The transformations among them can be realized by tuning the setting angles of the half-wave plates (HWPa1, HWPa2, HWPb1, HWPb2, HWPb3). 

\subsection{Projective Measurement and Detection}

As discussed in\,\cite{gupta2023quantum}, we use a 5-cycle graph to represent the KCBS contextually scenario. The five vertices stand for a set of these five observables as projectors in a three-dimensional Hilbert space. The edge connecting two vertices corresponds to two projectors that are mutually orthogonal to each other and can be measured jointly. We extend the 5-cycle graph by adding additional vertices $6, 7, 8$, assigning additional projectors, to ensure each vertex belongs to a complete orthogonal basis of dimension three. 

\textit{Measurement Strategy}---
To perform projective measurements in a three-dimensional (qutrit) Hilbert space, we implement a measurement setup that discriminates among the three mutually orthogonal basis states $\{ |\phi_i\rangle, |\phi_j\rangle,|\phi_k\rangle \}$ corresponding to a particular measurement setting. Each state $ |\phi_i\rangle$ is transformed via a unitary operation into one of the encoding basis states $\{ |0\rangle, |1\rangle,|2\rangle\}$ using wave plates and beam displacers. After the transformation, the three encoding basis states are routed to three spatially separated detectors, $D_1, D_2, D_3$, respectively.

\begin{figure}[h!]
    \centering
\includegraphics[width=0.65\linewidth]{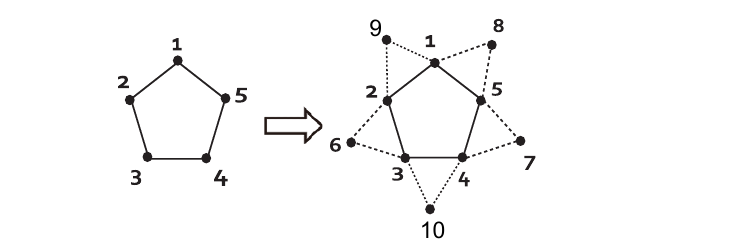}
    \caption{The construction of the extended graph from the 5-cycle graph. }
    \label{figs2}
\end{figure}

\textit{Binary Output Convention}---
Although each projective measurement technically has three possible outcomes, in our contextuality-based QKD protocol, we assign a binary result to each measurement. Specifically, each measurement setting is associated with a designated "preferred" projector—typically the one aligned with the contextuality inequality under test  (e.g., projector $\Pi_i = |\phi_i\rangle \langle \phi_i|$). A detection event in the corresponding detector (say $D_1$) is assigned the outcome $0$, while detections in the other two detectors are treated as $1$ outcomes. 
Each measurement setting is configured by adjusting the angles of the half-wave plates (HWPb1, HWPb2, HWPb3) in Bob's module. These angles are pre-calculated to implement the required unitary transformation for each measurement basis. In the current proof-of-principle demonstration, all measurement bases are selected and aligned manually for stability and precision.
This allows us to construct binary-valued observables suitable for evaluating the contextuality inequality (e.g., the KCBS expression).

\subsection{The dimension witness}

The dimension witness tested here is formulated within a prepare-and-measure scenario and is inspired by \cite{gupta2023quantum}. The witness involves nine possible preparations, denoted by $x=0,1,\cdots,8$ and eight possible binary-outcome measurements, denoted by $y=1,\cdots,8$. The measurement outcomes are labelled as $z=0,1$. In each experimental run, $x$ and $y$ are chosen randomly, and $p(z|x,y)$ represents the probability of obtaining outcome $z$ given that preparation $x$ and measurement $y$. The witness is constructed based on the extended KCBS graph (shown in Fig. \ref{figs2}) and is explicitly given by:
\begin{equation}\label{fomk}
S = \frac{1}{35} \left[\underbrace{\sum_{x=1}^{8}p(0|x,y=x) + \sum_{x=1}^{8} \sum_{y\in N_x} p(1|x,y)}_{S_1} + \underbrace{\sum_{y=1}^5 p(0|0,y)}_{S_2}\right].
\end{equation}
The witness consists of two terms, $S_1$ and $S_2$. The $S_1$ term enforces that the outcome $z$ should be 0 when the inputs $x$ and $y$ are identical, and the outcome should be 1 when the inputs $x$ and $y$ are neighbours (or connected by an edge) in the extended graph. Here, $N_x$ denotes the set of vertices that are connected to $x$ by an edge in the graph. The second term, $S_2,$ corresponds to the KCBS noncontextuality inequality involving preparation $x=0$. We assume that the dimension of the system is at most three. The optimal classical value of this witness is $S_c = 32/35$ and the quantum value using the KCBS states $S_Q = (30+\sqrt{2})/35$. The experimentally observed values are $S_1=29.8238\pm 0.0129$ and $S_2=2.2463\pm0.0091$, such that $S=S_1+S_2=32.0701\pm 0.0221$.

\subsection{Experimental Data}
In this section, we present the experimental correlations of all the experimental data that are relevant to the tests of the KCBS inequality ($S_2$) and the orthogonalities among all the bases ($S_1$) for two different photon sources.

For the single-photon source, the time-tagging module is configured with a time window of 13ns, corresponding to the repetition rate of the resonant excitation laser. A total of 4,000,000 time bins are recorded to ensure high statistical confidence.

For the weak coherent laser source, we perform measurements using varying time windows of 100 ns, 200 ns, ..., 1000 ns, and then 1500 ns, 2000 ns, ..., up to 4000 ns, with 1,000,000 bins recorded for each setting. This allows us to simulate different average photon numbers and analyze their impact on the violation of the KCBS inequality.

\begin{table}[ht]
\centering
\begin{tabularx}{0.7\textwidth}{c|>{\centering\arraybackslash}X>{\centering\arraybackslash}X>{\centering\arraybackslash}X}
\toprule
\toprule
 & \multicolumn{3}{c}{measure(y)} \\
prepare(x) & $y=1$ & $y=2$ & $y=9$ \\
\midrule
$x=1$ & 0.989465 & 0.005566 & 0.00497 \\
$x=2$ & 0.006559 & 0.991230 & 0.002211 \\
$x=9$ & 0.005838 & 0.006355 & 0.987808 \\
\midrule
 & \multicolumn{3}{c}{measure(y)} \\
prepare(x) & $y=1$ & $y=5$ & $y=8$ \\
\midrule
$x=1$ & 0.973752 & 0.016821 & 0.009427 \\
$x=5$ & 0.004526 & 0.988657 & 0.006817 \\
$x=8$ & 0.007934 & 0.005435 & 0.986632 \\
\midrule
 & \multicolumn{3}{c}{measure(y)} \\
prepare(x) & $y=3$ & $y=4$ & $y=10$ \\
\midrule
$x=3$  & 0.994900 & 0.003466 & 0.001635 \\
$x=4$  & 0.000786 & 0.996641 & 0.002573 \\
$x=10$ & 0.004333 & 0.006579 & 0.989088 \\
\midrule
 & \multicolumn{3}{c}{measure(y)} \\
prepare(x) & $y=5$ & $y=4$ & $y=7$ \\
\midrule
$x=5$ & 0.992029 & 0.004405 & 0.003566 \\
$x=4$ & 0.007522 & 0.983276 & 0.009201 \\
$x=7$ & 0.004324 & 0.001580 & 0.994096 \\
\midrule
 & \multicolumn{3}{c}{measure(y)} \\
prepare(x) & $y=3$ & $y=2$ & $y=6$ \\
\midrule
$x=3$ & 0.989063 & 0.001967 & 0.00897 \\
$x=2$ & 0.001481 & 0.996499 & 0.00202 \\
$x=6$ & 0.001198 & 0.003194 & 0.995608 \\
\bottomrule
\bottomrule
\end{tabularx}
\caption{Experimental data that evaluates the orthogonality, i.e., the binary projective measurements for different prepare(x) and measure(y) that constitute $S_1$.}
\end{table}

\begin{table}[ht]
\centering
\begin{tabularx}{0.7\textwidth}{c|>{\centering\arraybackslash}X>{\centering\arraybackslash}X>{\centering\arraybackslash}X}
\toprule
\toprule
 & \multicolumn{3}{c}{measure(y)} \\
prepare(x) & $y=1$ & $y=2$ & $y=9$ \\
\midrule
$x=0$ & 0.453846 & 0.448252 & 0.097902 \\
\midrule
 & $y=1$ & $y=5$ & $y=8$ \\
\midrule
$x=0$ & 0.455885 & 0.455334 & 0.08878 \\
\midrule
 & $y=3$ & $y=4$ & $y=10$ \\
\midrule
$x=0$  & 0.451435 & 0.4469 & 0.101666 \\
\midrule
  & $y=5$ & $y=4$ & $y=7$ \\
\midrule
$x=0$ & 0.44229 & 0.458833 & 0.098877 \\
\midrule
  & $y=3$ & $y=2$ & $y=6$ \\
\midrule
$x=0$ & 0.452994 & 0.454653 & 0.092353 \\
\bottomrule
\bottomrule
\end{tabularx}
\caption{Experimental data that evaluates the KCBS inequality, i.e., the binary projective measurement outcomes for different measures (y) while keeping the same initial state (x=0) that constitute $S_2$.}
\end{table}

\begin{table}[ht]
\centering
\begin{tabularx}{\textwidth}{c|>{\centering\arraybackslash}X>{\centering\arraybackslash}X>{\centering\arraybackslash}X>{\centering\arraybackslash}X>{\centering\arraybackslash}X>{\centering\arraybackslash}X>{\centering\arraybackslash}X>{\centering\arraybackslash}X}
\toprule
\toprule
Average photon number & 0.009117 & 0.006612 & 0.005325 & 0.004645 & 0.004149 & 0.003747 & 0.003486 & 0.003269 \\
\midrule
KCBS inequality & 2.24741 & 2.23101 & 2.19124 & 2.15935 & 2.13154 & 2.10740 & 2.08223 & 2.05275 \\
\midrule
Average photon number & 0.003048 & 0.002889 & 0.002309 & 0.002001 & 0.001784 & 0.001628 & 0.001490 & 0.001389 \\
\midrule
KCBS inequality & 2.02786 & 2.00627 & 1.88515 & 1.77194 & 1.67118 & 1.59101 & 1.49143 & 1.42113 \\
\bottomrule
\bottomrule
\end{tabularx}
\caption{Experimental values of the KCBS inequality violation for a weak coherent laser source as a function of the average photon number. The average photon number is adjusted by varying the detection time window, and the corresponding KCBS values are calculated from the observed measurement correlations. As the average photon number increases, multi-photon events become more likely, leading to a monotonic decrease in the KCBS inequality violation.}
\end{table}

\clearpage

 \section{\texorpdfstring{Theoretical: key analysis for the experimental results}{key analysis for the experimental results for S1 and S2}}

\subsection{Application for Quantum Key Distribution}

After completing a large number of experimental runs, they randomly select a subset of these runs and publicly disclose their inputs and outcomes to compute the value of the dimension witness $S$. For the remaining runs, used for key generation, Bob announces his measurement input $y$. Alice then announces to Bob to discard the run if her preparation input was $x=0$. Since Alice knows both $x$ and $y$, she can perfectly predict Bob's outcome in the retained runs, which serves as the raw key. Therefore, the key is 0 if $y=x$, and the key is 1 if $y\in N_x$. Note that key generation is restricted to the runs when $y \in \{N_x,x\}$ (or the runs that appear in $S_1$), ensuring that Alice and Bob have the perfectly correlated key. However, both $S_1$ and $S_2$ are utilized for security verification.

We calculate the probability ($P_k$) that a particular run of the task contributes to key generation. Let Alice's and Bob's inputs be independently and uniformly distributed, that is, $p(x)=1/9$ and $p(y)=1/8$. Then, the probability that $y\in \{N_x,x\}$ according to the extended graph (Figure 1 of \cite{gupta2023quantum}) is $30/72$. These runs are included in $S_1$. Among them, half of the runs (randomly selected by the users) are used to evaluate $S$ for verification, while the remaining half contribute to key generation. Therefore, the probability that a run is used for key generation is $P_k = 30/144 \approx 0.208.$

We can apply Theorem 2 of \cite{gupta2023quantum} to conclude that the states sent by Alice's $\{\rho_x\}$ are pure states and satisfy the orthogonality relations according to the 5-cycle graph, since $S_1=30$. Here, Bob receives pure states $\rho_x$ for $x=1,\cdots,8$, which leads to two possible cases.  

If the states $\rho_x$ ($x=1,\cdots,8,$) are diagonal states in some basis, then without loss of generality, they can be taken from the set $\{\ket{0},\ket{1},\ket{2}\}$. In this case,  since the states are orthogonal, Eve could measure in that basis or even make copies of the states, with some non-zero probability. 
However, in this case, $\rho_1$ and $\rho_3$ or $\rho_1$ and $\rho_4$ have to be orthogonal, and similarly for other pairs of states, implying contextuality cannot be observed. 
Yet, the users obtain $S_2>2$, which contradicts the possibility that Eve performs any quantum operation. 

If the states $\rho_x$ ($x=1,\cdots,8,$)  are not diagonal states, which means that at least two of the states are superposition of $\{\ket{0},\ket{1},\ket{2}\}$, to maintain the orthogonality relations of the graph. Any quantum operation (apart from applying a unitary) performed by Eve with some non-zero probability on $\rho_x$ would change the orthogonality relations, making it impossible to achieve $S_1=30$. Thus, Eve cannot be present.

As a consequence, the best strategy available for Eve is to guess the $f(x,y)$ for $y \in \{N_x,x\}$, which is optimally attained when Eve always guesses the key to be 1. Therefore, the key rate is, 
\begin{eqnarray}
    r &=& I(A:B) - I(A:E) \nonumber \\
    &=& \sum_{a,b=0,1} p(z=a,f=b)\frac{p(z=a,f=b)}{p(z=a)p(f=b)} - \sum_{a,b=0,1} p(e=a,f=b)\frac{p(e=a,f=b)}{p(e=a)p(f=b)} \nonumber  \\
    & =& 0.8366, 
\end{eqnarray}
where $p(z=a),p(f=a),p(e=a)$ denote the probability that Bob's outcome $z=a$, the function $f(x,y)=a$, Eve's outcome $e=a$, respectively. Thus, the overall key rate is $P_k\cdot 0.8366 = 0.174.$

\subsubsection{\texorpdfstring{The case when $S = S_1+S_2 > S_c = 32$ and $S_1<30$.}{The case when S = S1+S2 > Sc = 32 and S1 < 30}}

We consider an individual attack strategy, where Eve applies an arbitrary quantum channel to the state communicated by Alice. Let the post-channel states be denoted by $\{\tilde{\rho}_x\}_x$. In the QKD protocol, Bob publicly announces his input $y$, which Eve can access. Based on this information, Eve performs binary-outcome measurements $\{E_{e|y}\}$ with $e\in \{0,1\}$ to guess $f(x,y)$. Our goal is to determine the optimal average guessing probability for Eve under the constraint $S_1+S_2=32.0701.$ First, we use the semi-definite programming (SDP) relaxation technique based on the Navascu\'es--Pironio--Ac\'in (NPA) hierarchy \cite{npa}, which provides an upper bound on Eve’s guessing probability for $f(x,y)$.  This optimization is carried out over all possible quantum realizations of states $\{\tilde{\rho}_x\}$, Eve's measurements $\{E_{e|y}\}$, and Bob's measurements $\{M_{z|y}\}$ without assuming their Hilbert space dimensions. Formally, the problem is expressed as:
\begin{eqnarray}
  & \sup\limits_{\{E_{e|y}\},\{\tilde{\rho}_x\},\{M_{z|y}\}}  &  \frac{1}{30}\left( \sum_{x=1}^8 \tr(\tilde{\rho}_x E_{0|y}) + \sum_{x=1}^{8} \sum_{y\in N_x} \tr(\tilde{\rho}_x E_{1|y}) \right) \nonumber \\
  & \text{s.t.} & \tilde{\rho}_x \geq 0, \tr(\tilde{\rho}_x) = 1; \  M_{z|y} \geq 0, \sum_z M_{z|y} = \ I, \forall z,y;  \  E_{e|y} \geq 0, \sum_e E_{e|y} = \ I, \forall e,y. \nonumber \\
  && \sum_{x=1}^8 \tr(\tilde{\rho}_xM_{0|x}) + \sum_{x=1}^{8} \sum_{y\in N_x} \tr(\tilde{\rho}_x M_{1|y}) + \sum_{y=1}^5 \tr(\tilde{\rho}_0 M_{0|y}) \geq 32.0701, \nonumber \\
  && [M_{z|y},E_{e|y'}] = 0, \forall z,e,y,y'.
\end{eqnarray}
The constraint that Eve's and Bob’s measurements commute reflects their physically separated laboratories. Solving the SDP yields an upper bound of 1 on the guessing probability, which corresponds to a lower bound of zero on the key rate. Consequently, this approach does not provide a meaningful or nontrivial security bound. 

\begin{figure}[h!]
    \centering
\includegraphics[width=0.43\linewidth]{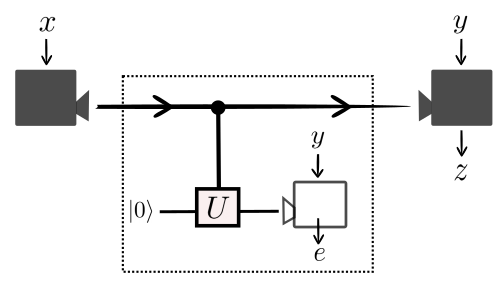}
    \caption{A natural and effective attack by Eve, illustrated in the dotted box, that involves copying the input value $x$ onto her ancilla, which she can later measure in some basis that depends on $y$.}
    \label{fig:abe}
\end{figure}

\textit{Security under a natural strategy of Eve.---} Next, we examine a natural and structured attack strategy. Eve possesses an ancilla initialized in the qutrit state $\ket{0}$ and performs a controlled unitary gate that clones the communicated state if it belongs to the particular basis, say $\{\ket{0},\ket{1},\ket{2}\}$. Specifically, the controlled operation acts as:
\begin{equation}\label{eve-U}
    U\ket{0}\ket{0} = \ket{0}\ket{0}, \  U\ket{1}\ket{0} = \ket{1}\ket{1}, \ U\ket{2}\ket{0} = \ket{2}\ket{2} .
\end{equation}
This allows Eve to encode information about $x$ onto her ancilla and later measure it in a basis dependent on $y$. Since the secret key is derived from inputs $x\in \{1,\cdots,8\}$, this strategy is particularly effective. However, this operation alters the states received by Bob, rendering them diagonal in the computational basis. Consequently, the maximal value of $S$ cannot be larger than $S_c$. Therefore, Eve cannot perform this attack unconditionally. Instead, she applies it with some probability $q$, and with probability $(1-q)$, allows the state to pass unaltered (see Figure \ref{fig:abe}). 

\begin{figure}
    \centering
    \includegraphics[width=0.5\linewidth]{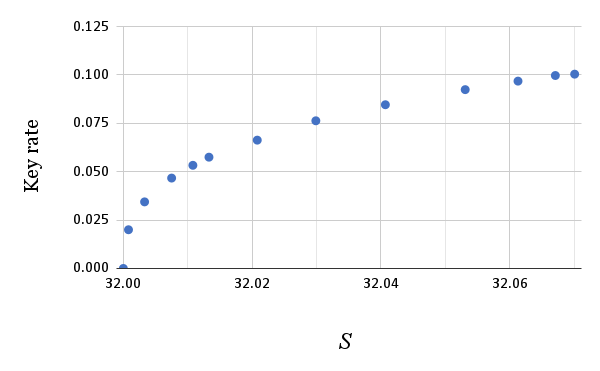}
    \caption{The overall key rate with respect to the value of $S$ is shown.}
    \label{fig:key-S}
\end{figure}

\begin{result}
    If Eve performs the generic attack described above, then the overall key rate with respect to the value of $S$ is shown in Figure \ref{fig:key-S}. In particular, the overall key rate is 0.1004 when $S = 32.0701$ is observed. 
\end{result}
To compute the guaranteed key rate, we optimize Eve's guessing probability of $f(x,y)$, while fixing her aforementioned generic strategy. The optimization is done in the following way. We fix a value of the probability $q$ from the set $\{0,1\}$ starting from 1 and then decreasing in the interval of 0.01. Then we optimize the expression of $S$ where $\{\rho_x\}$ and $\{M_{z|y}\}$ are unknown but act on $\mathbb{C}^3$. The optimization is done under the condition that whenever Eve performs the attack, she is perfectly able to guess $f(x,y)$. Moreover, we consider all possible combinations so that the measurement operators $\{E_{0|y}\}$ belong to the set $\{\ket{0}\!\bra{0},\ket{1}\!\bra{1},
\ket{2}\!\bra{2}\}$ and the unitary $U$ is given by \eqref{eve-U}. Formally, we execute the following optimization:
\begin{eqnarray}
  & \sup\limits_{\{E_{e|y}\},\{M_{z|y}\},\{\rho_x\}}  & q \bigg(\sum_{x=1}^8 \tr(U(\rho_x \otimes \ket{0}\!\bra{0})U^\dagger(M_{0|x}\otimes \I)) + \sum_{x=1}^{8} \sum_{y\in N_x} \tr(U(\rho_x \otimes \ket{0}\!\bra{0})U^\dagger(M_{1|y}\otimes \I)) \nonumber \\
   &&  + \sum_{y=1}^5 \tr(U(\rho_0\otimes \ket{0}\!\bra{0})U^\dagger(M_{0|y}\otimes \I)) \bigg) + (1-q) \bigg(\sum_{x=1}^8 \tr(\rho_x M_{0|x}) + \sum_{x=1}^{8} \sum_{y\in N_x} \tr(\rho_xM_{1|y})  + \sum_{y=1}^5 \tr(\rho_0M_{0|y}) \bigg) \nonumber \\
  & \text{s.t.} & \rho_x \geq 0, \tr(\rho_x) = 1, \forall x; \ M_{z|y} \geq 0, \sum_z M_{z|y} = \I, \forall z,y; \ M_{z|y}, \rho_x \text{ acts on } \mathbb{C}^3; \nonumber \\
   && \sum_{x=1}^8 \tr(U(\rho_x \otimes \ket{0}\!\bra{0})U^\dagger(\I \otimes E_{0|y})) + \sum_{x=1}^{8} \sum_{y\in N_x} \tr(U(\rho_x \otimes \ket{0}\!\bra{0})U^\dagger(\I \otimes E_{1|y})) = 30.
\end{eqnarray}
This optimization is executed using a SeeSaw approach: we iteratively fix the set of states $\{\rho_x\}$ and optimize the value of $S$ over the measurements $\{M_{z|y}\}$, then fix the optimized measurements and re-optimize $S$ over the states $\{\rho_x\}$. We repeat this process until convergence. We note down the resultant maximum value of $S$ (which is less than or equal to 32.0701) along with the quantum strategy and subsequently calculate the overall key rate. The explicit example when $S=32.0701$ is given below. Then we decrement $q$ and repeat the procedure.

It is obtained that at $q=0.54,$ the maximum value of $S$ is 32.0701. The quantum strategy that achieves this is given as follows. Alice's device sends the following states for different inputs $x$,
\begin{equation}
    \rho_1 = \rho_3= \rho_7 = \ket{0}\!\bra{0}, \rho_2 = \rho_4 = \rho_8 = \ket{1}\!\bra{1},\rho_5 = \rho_6 = \ket{2}\!\bra{2}, \rho_0 = \ket{\psi}\!\bra{\psi}, \ \ket{\psi} = \frac{1}{\sqrt{2}}\left( \ket{0} + \ket{1} \right)  .
\end{equation}
The measurements performed in Bob's device are given by,
\begin{eqnarray*}
M_1 = M_3 =
\begin{bmatrix}
0.9932 & -0.0822 &  0 \\
-0.0822 &  0.0068 & 0 \\
0 & 0 & 0
\end{bmatrix} , 
M_2 = M_4 = 
\begin{bmatrix}
0.0068  & -0.0822 &  0 \\
-0.0822  &  0.9932 & 0 \\
0 & 0 & 0
\end{bmatrix} , M_5 = M_6 = \ket{2}\!\bra{2}, M_7 = \ket{0}\!\bra{0}, M_8 = \ket{1}\!\bra{1} .
\end{eqnarray*}
Eve's measurements are $E_{0|y} = \rho_y$ for $y=1,\cdots,8.$ In this attack, Eve perfectly learns $f(x,y)$ with probability $q=0.54$, and with probability $(1-q)=0.46$, she makes the best possible guess. The corresponding key rate is:
\begin{eqnarray}
    r &=& I(A:B) - I(A:E) \nonumber \\
    &=& \sum_{a,b=0,1} p(z=a,f=b)\frac{p(z=a,f=b)}{p(z=a)p(f=b)} - \sum_{a,b=0,1} p(e=a,f=b)\frac{p(e=a,f=b)}{p(e=a)p(f=b)} \nonumber  \\
    & =& 0.8109 - 0.3292 = 0.4817, 
\end{eqnarray}
Thus, the overall key rate is $P_k\cdot 0.4817 = 0.1004.$

\subsection{Application for Quantum randomness generation} 

As discussed in the Supplementary Material of \cite{gupta2023quantum}, the random numbers are generated from the experimental runs where $x=0$. The amount of certified randomness is quantified by
\begin{equation}
    R = - \log_2 p^*, \quad \text{ where } p^* = \max_{z,y} \{p(z|x=0,y)\} .
\end{equation}
An upper bound on $p^*$ can be obtained by solving the following optimization problem.
\begin{eqnarray} \label{opt:randomness}
  & \sup\limits_{\{\rho_x\},\{M_{z|y}\}}  & \max_{y=1,\cdots,5} \left\{ \tr\left(\rho_0 M_{0|y} \right), 1-\tr\left(\rho_0 M_{0|y} \right)\right\} \nonumber \\
  & \text{s.t.} & \rho_x \geq 0, \tr(\rho_x) = 1,  \ \rho_x \text{ acts on } \mathbb{C}^3, \forall x, \nonumber \\
  & & M_{z|y} \geq 0, \sum_z M_{z|y} = \I,  \  M_{z|y} \text{ acts on } \mathbb{C}^3, \forall z,y, \nonumber \\
  && \sum_{x=1}^8 \tr(\rho_x M_{0|x})+  \sum_{x=1}^{8} \sum_{y\in N_x} \tr(\rho_xM_{1|y}) \geq S_1, \nonumber \\
  && \sum_{y=1}^5 \tr(\rho_0 M_{0|y}) \geq S_2. 
\end{eqnarray}
Given that the states $\{\rho_x\}$ are defined over the Hilbert space $\mathbb{C}^3$, we apply the semi-definite relaxation technique developed by Navascués and Vértesi for bounding quantum correlations in finite dimensions \cite{NVHierarchy}.

\begin{result}
Performing the optimization \eqref{opt:randomness} by the semi-definite relaxation proposed in \cite{NVHierarchy}, with the value $S_1 = 29.8238$ and $S_2 = 2.2463,$ we find that the upper bound of $p^*$ is 1, which implies no certifiable randomness ($R=0$). However, when the values are slightly adjusted to $S_1=30$ and $S_2 = 2.2463,$ the upper bound on $p^*$ reduces to $0.5510$, corresponding to a certified randomness of $R=0.86$ bits.
\end{result}

Finally, we present an analytical result showing that when $S_1=30$, the generated randomness approaches its maximum as $S_2$ approaches its maximal value of $\sqrt{5}$.

\begin{result}
 When $S_1=30,$ for any observed value of $S_2 \geqslant  (\sqrt{2} - \epsilon)$, 
 \begin{equation}\label{p*ep}
    p^* =  \max_{z,y} \{p(z|x=0,y)\} 
     \leq 1-1/\sqrt{5} + 2\mathcal{O}(\sqrt{\epsilon}) .
 \end{equation}
 Therefore, $R = -\log_2 (0.553 + 2\mathcal{O}(\sqrt{\epsilon})).$
\end{result}

\begin{proof}
If $S_1=30$, we can invoke Theorem 2 from \cite{gupta2023quantum} to conclude that the states sent by Alice's $\{\rho_x\}$ are pure states and satisfy the orthogonality relations according to the 5-cycle graph. Consequently, we can apply the self-testing result from \cite{bharti_prl}, which implies there exists a unitary $U$, such that
\begin{equation}  \label{st}  
||U\rho_0 U^\dagger - |0\rangle\!\langle 0| || \leqslant \mathcal{O}(\sqrt{\epsilon}) , \ ||UM_{0|x} U^\dagger - |\psi_x\rangle\!\langle \psi_x| || \leqslant \mathcal{O}(\sqrt{\epsilon}) ,
\end{equation}
where $x=1,2,3,4,5$, $\{|\psi_x\rangle\}$ are the optimal KCBS states, and $\rho_0$ and $\{M_{0|x}\}$ are the unknown state and measurements. 
Let us define $\rho'_0 = U\rho_0U^\dagger$ and $M'_{0|y} = UM_{0|y}U^\dagger$. Using the self-testing relation \eqref{st}, along with some operator identities, we establish that
\begin{eqnarray}
    |p(0|0,y) - 1/\sqrt{5} | & = & |\tr(\rho'_0 M'_{0|y} ) - \tr(|0\rangle\!\langle 0||\psi_x\rangle\!\langle \psi_x|)| \nonumber \\
    & = & || \rho'_0 M'_{0|y} - |0\rangle\!\langle 0||\psi_x\rangle\!\langle \psi_x| || \nonumber \\
    & \leq & || \rho'_0 M'_{0|y} - \rho'_0|\psi_x\rangle\!\langle \psi_x| || + ||\rho'_0|\psi_x\rangle\!\langle \psi_x| - |0\rangle\!\langle 0||\psi_x\rangle\!\langle \psi_x| || \nonumber  \\
    & \leq & || M'_{0|y} -|\psi_x\rangle\!\langle \psi_x| || + ||\rho'_0 - |0\rangle\!\langle 0| || \nonumber  \\\
    & \leq & 2\mathcal{O}(\sqrt{\epsilon}).
\end{eqnarray}
Therefore, 
\begin{equation}
    1/\sqrt{5} - 2\mathcal{O}(\sqrt{\epsilon}) \leq p(0|0,y) \leq 1/\sqrt{5} + 2\mathcal{O}(\sqrt{\epsilon}),
\end{equation}
which implies that the minimum value of $p(0|0,y)$ is $1/\sqrt{5} - 2\mathcal{O}(\sqrt{\epsilon})$. Thus, the maximum value of $p(1|0,y)$ is given by the right-hand-side of \eqref{p*ep}. 
\end{proof}

\end{document}